# Electronic sideband locking of 318.6nm UV laser to an ultrastable optical cavity with a wide continuously tunable range

Jiandong Bai,[1,2] Jieying Wang,[1,2] Jun He,[1,2,3] and Junmin Wang [1,2,3,*]

[1] *Institute of Opto-Electronics, Shanxi University,*

*Tai Yuan 030006, Shan Xi, China*

[2] *State Key Laboratory of Quantum Optics and Quantum Optics Devices, Shanxi University,*

*Tai Yuan 030006, Shan Xi, China*

[3] *Collaborative Innovation Center of Extreme Optics, Shanxi University,*

*Tai Yuan 030006, Shan Xi, China*

* *wwjjmm@sxu.edu.cn*

**Abstract:** We have demonstrated a frequency-stabilized tunable 318.6 nm ultraviolet (UV) laser system for the single-photon $6S_{1/2}$ - nP (n = 70 ~ 100) Rydberg excitation of cesium atoms. The 637.2 nm laser produced by single-pass sum frequency generation from two infrared fiber lasers is offset locked to a high-finesse ultra-low expansion (ULE) optical cavity placed in ultra-high vacuum using the electronic sideband locking technique. The generated 318.6 nm UV laser via cavity-enhanced second-harmonic generation can be continuously tuned over 4 GHz by indirectly changing modulation frequency on the electro-optic phase modulator while the whole laser system remains locked. We analyze the tuning range mainly depends on the modulator bandwidth and the tunable range of the seed laser. The locking scheme offers a method to compensate the frequency difference between the reference frequency and the goal frequency to a desired excited state, and has huge potential in precision spectroscopic experiments of cold atoms.

## 1. Introduction

Considering the Rydberg blockade mechanism introduced by the strong long-range dipole-dipole interaction between highly excited Rydberg atoms [1-5], Rydberg states of neutral atoms have great potential in practical applications, such as realization of the quantum logic gate [6], quantum entanglement between two atoms [7, 8], quantum communication [9] and quantum information processing [1]. For high Rydberg states, because of the low transition probability and the narrow transition linewidth, people usually choose a two-step or three-step excitation to a specific Rydberg state. However, the multi-step excitation will inevitably appear the population of intermediate state, which brings photon scattering and AC-Stark shift resulting in the low excitation efficiency. The corresponding disadvantages can be avoided by single-photon excitation. Recently, Tong *et al.* [10] and Hankin *et al.* [11] demonstrated direct Rydberg excitation of $^{85}$Rb at 297 nm and $^{133}$Cs at 319 nm by single-photon transition, respectively. For single-photon Rydberg excitation of cesium atoms from $6S_{1/2}$ to nP (n =70 ~ 100) state, the 318.6 nm UV laser should be continuously tuned over a wide range to address the desired Rydberg state. Due to the energy level spacing between adjacent Rydberg states is relatively dense, the UV laser should be narrow linewidth and frequency stabilization. To produce the 318.6 nm UV laser, the frequency doubling requires a frequency-stabilized 637.2 nm laser with a tunable central frequency as the fundamental wave.

To improve the laser frequency stability, it is crucial to select a frequency reference standard, such as atomic or molecular transition spectral lines, resonance lines of the stable optical cavity. Compared with atomic or molecular transition spectral lines, a Fabry-Pérot (F-P) cavity made of ultralow expansion (ULE) glass placed in an ultra-high vacuum has high stability and ultra-narrow resonance linewidth, and can almost realize frequency stabilization of all lasers. To obtain a frequency-stabilized tunable light source, a common method is locking a laser to a fixed optical cavity, and then inserts one or more acousto-optic modulators (AOM) in optical path [12]. AOMs are commonly used for MHz-level shifts, but a given AOM has a limited tuning range that is only a small fraction of its fixed center frequency. By double-passing AOM with a curved mirror or the cat-eye configuration, a frequency-dependent deflection of the light beam which is produced by AOMs can be mitigated. AOMs used for GHz-level shifts offer this increased bandwidth by sacrificing diffraction efficiency, and they are very expensive. Broadband electro-optic phase modulators (EOPMs) avoid many of the bandwidth and tuning range limitations of AOMs. Recently, Thorpe *et al.* [13] proposed the sideband locking technique which is a few modifications for the standard Pound Drever - Hall (PDH) [14] locking method to obtain a stably tunable laser. Subsequently, Johnson *et al.* [15] and Kohlhaas *et al.* [16] reported the relative frequency stabilization of a laser on an optical cavity by serrodyne frequency shifting. Serrodyne modulation was applied by feeding the output of a nonlinear transmission line into an EOPM. By changing the frequency of the sinusoidal wave at the input of the nonlinear transmission line, the frequency of the light at the output of the EOPM is tuned.

In this paper, we use the electronic sideband locking (ESB) scheme to produce a frequency-stabilized laser with a tunable central offset frequency. Generally, a commercial EOPM has limited RF bandwidth. For example, the maximum RF bandwidth of a bulk-EOPM with the electro-optical crystal of $LiNbO_3$ from New Focus is about 250 MHz. Compared with the bulk-EOPM, the waveguide-type EOPM has a larger RF bandwidth and a smaller RF power consumption. For example, the waveguide-type EOPMs from Eospace, Photoline, and Jenoptik normally have a typical bandwidth of up to dozens of GHz. We apply PDH and ESB technique for the frequency stabilization of two distributed feedback fiber lasers at 1560.5 nm and 1076.9 nm to a ULE reference cavity. We follow the approach demonstrated by David Wineland's group in their pioneered work [17] to generate the desired 318.6 nm UV laser starting from two infrared (IR) fiber-laser sources. With a mode-matched incident power of about 2.35 W, we achieve 1.34 W of UV laser output, corresponding to a conversion efficiency of 57%. Based on a high-finesse ULE cavity and the waveguide-type EOPM driven by a phase-modulated sinusoidal wave signal, we realize a stable high-power tunable 318.6 nm UV laser by changing modulation sideband on EOPM. The laser system can be used for the future single-photon Rydberg excitation of cesium atoms from $6S_{1/2}$ to nP (n = 70 ~ 100).

**2. Experimental setup**

A schematic of the experimental setup is shown in Fig. 1. Two infrared seed lasers are single-frequency 1560.5 nm distributed feedback Erbium-doped fiber laser (DFB-EDFL) and 1076.9 nm distributed feedback Ytterbium-doped fiber laser (DFB-YDFL) from NKT Photonics with nominal maximum output power of 200 mW and 100 mW, respectively. Two fiber lasers seed commercial 15 W Erbium-doped fiber amplifier (EDFA) and 10 W Ytterbium-doped fiber amplifier (YDFA), respectively. A fiber-based broadband waveguide-type EOPM-1 from Eospace is placed between 1560.5 nm DFB-EDFL and 15 W EDFA to produce a set of sidebands (modulation frequency: 12.6 MHz). The 1560.5 nm seed laser is separated off for the doubling cavity locking, and for locking to a

ULE optical cavity. The boosted lasers are combined and passed through a 40mm × 10mm × 0.5mm MgO-doped periodically poled lithium niobate (PPMgO:LN) crystal with a poling period of 11.80 μm. In our previous work [18], the crystal is temperature stabilized at 154.0 °C to produce 8.75 W of 637.2 nm red laser by single-pass sum frequency generation (SFG) process. The generated 637.2 nm laser is separated by a half-wave plate (HWP) and polarization beam splitter (PBS) cube. A small part of the 637.2 nm laser (~3 mW) is phase modulated by a fiber-coupled waveguide-type broadband EOPM-2 from Jenoptik (PM635), and then is locked to the ULE cavity to stabilize the frequency of 1076.9 nm DFB-YDFL using the ESB locking technique. A main portion of its output is used to frequency doubling to 318.6 nm UV laser via the enhancement cavity within a Brewster-cut $\beta$-BaB$_2$O$_4$ (BBO) crystal with a size of 10 mm × 3mm × 3mm.

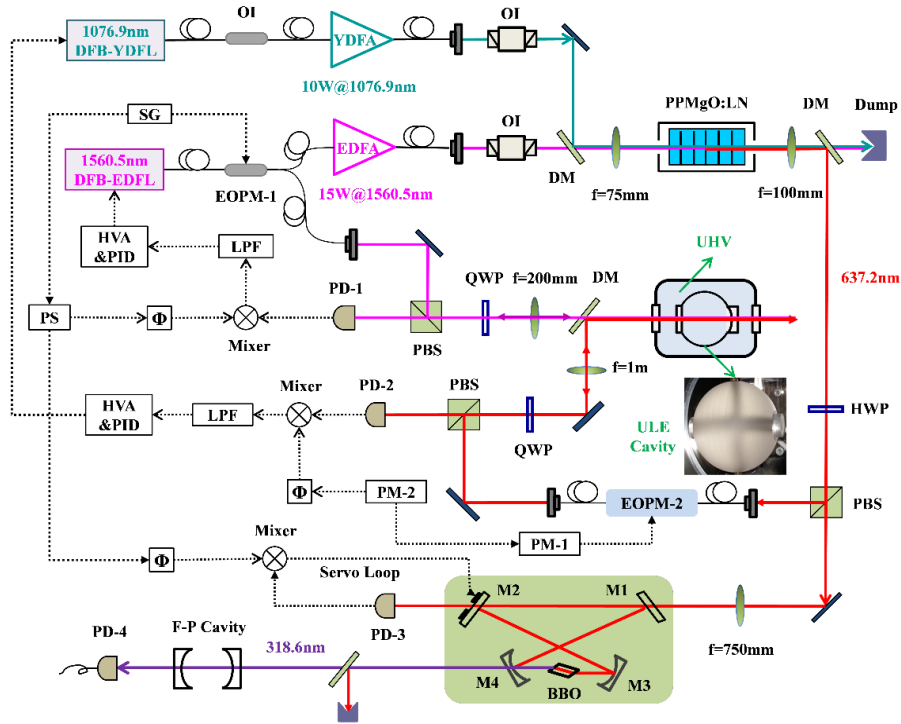

Fig. 1. Schematic of the experimental setup. Dotted curves and solid curves represent the circuitry part and the light path, respectively. DFB-YDFL, distributed feedback Ytterbium-doped fiber laser; YDFA, Ytterbium-doped fiber amplifier; DFB-EDFL, distributed feedback Erbium-doped fiber laser; EDFA, Erbium-doped fiber amplifier; OI, optical isolator; DM, dichromatic mirror; HWP, half-wave plate; QWP, quarter-wave plate; PBS, polarization beam splitter cube; EOPM, electro-optic phase modulator; SG, signal generator; PS, power splitter; HVA, high-voltage amplifier; PID, proportional-integral-differential controller; LPF, low-pass filter; Φ, phase shift; PD, photodiode detector; PM, phase modulation generator; ULE, ultralow expansion; UHV, ultra-high vacuum; M, 45° high-reflectivity mirror.

The ULE reference cavity from AT Films is a two-mirror spherical F-P cavity composed of plane-plane and plane-concave mirrors. The plane-concave mirror's radius of curvature is 500 mm. Both mirrors and the cavity body consist of ULE glass. The mirrors' reflective surfaces have high reflectivity at the operating wavelength of 1560.5 nm and 637.2 nm while another surfaces are anti-reflection coated for both wavelengths. The optical cavity is 47.6 mm long, and resulting in a 3.145 GHz free spectral range (FSR). The cavity design is similar to that described in [19, 20]. Two stainless steel screws hold the cavity at two points on a diameter of the spherical spacer that is orthogonal to the optical axis. Contact between the screws and the cavity spacer is made by two spherical vented Torlon, and the structure is designed such that the rigid-body motional modes of the sphere are at frequencies of a few hundred hertz. The optical cavity is housed in a thermal radiation shield inside a temperature-stabilized and ultra-high vacuum chamber to ensure a uniform and stable temperature. The temperature of the vacuum chamber is actively stabilized at 25 $^{o}$C. Typical pressure in the vacuum chamber is kept at ~ $3\times10^{-9}$ Torr by using an ion pump with a size of 15cm×12cm×8cm. By the modulation sideband method, the finesse of the ULE cavity is measured to be $3.4(2)\times10^4$ (FWHM ~ 92 kHz) at 1560.5 nm and $3.0(2)\times10^4$ (FWHM ~ 105 kHz) at 637.2 nm, respectively.

## 3. Results and discussion

A part of the phase-modulated 1560.5 nm laser via EOPM-1 is injected into the temperature-stabilized ULE cavity, and the response is monitored by PD-1 in reflected cavity signal. The sinusoidal modulation signal generated by a signal generator (SG) is divided into two parts to drive the EOPM-1 and enter into a power splitter (PS). One path through the PS is phase shifted, and then is mixed with the detected signal to generate the error signal. The error signal is amplified and optimized to stabilize the 1560.5 nm laser's frequency via servo feedback loop. Another part through the PS is combined with the detected signal from PD-3 to produce the error signal for locking doubling cavity. The 1560.5 nm laser frequency can be coarsely tuned ~145 GHz by slowly changing the temperature of the 1560.5 nm DFB-EDFL from 20℃ to 50℃. In the process of addressing different desired Rydberg states, the tunable 318.6nm laser is required. By changing the 1560.5 nm or 1076.9 nm lasers' frequency, we can realize the frequency tuning of the 318.6nm laser, which is doubling frequency from 637.2 nm laser. Here we concern that which the longitudinal mode of the 1560.5 nm laser, resonant with that of the ULE cavity, is closely relative to the desired Rydberg state. The resonance frequency of the 1560.5 nm laser is monitored by a wavemeter (Advantest, Q8326), as shown in Fig. 2(a). When the specific longitudinal mode of the 1560.5 nm laser is locked to the ULE cavity, a wide range of frequency tuning of the 637.2 nm laser can be achieved by changing the temperature of the 1076.9 nm DFB-YDFL with a temperature tuning range of ~202 GHz. Thus, the corresponding temperature coarse tuning of the UV laser is ~404 GHz. When the 1560.5 nm and 1076.9 nm lasers remain locked, the longitudinal mode of the 637.2 nm laser is also monitored by a wavemeter to roughly align a Rydberg transition line of the cesium atoms, as shown in Fig. 2(b).

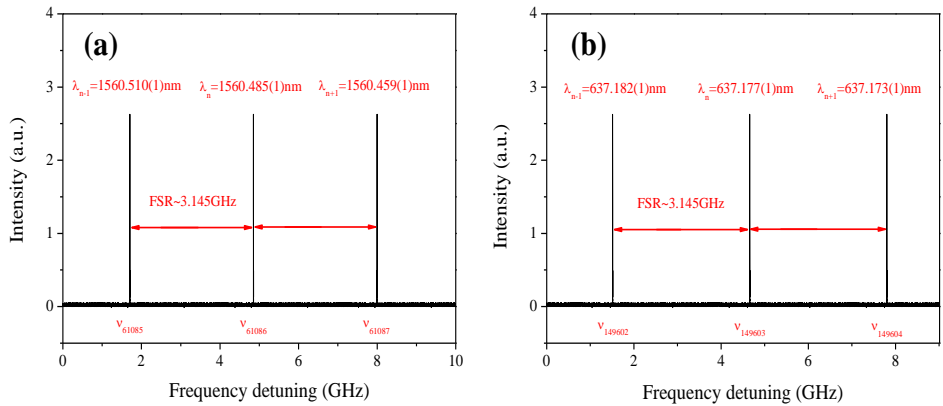

Fig. 2. (a) Frequency tuning of the 1560.5 nm laser by changing the temperature of EDFL; (b) Frequency tuning of the 637.2 nm laser by changing the 1076.9 nm laser's temperature when the 1560.5 nm laser is locked to the ULE cavity.

The waveguide-type EOPM-2 is driven by a phase-modulated signal. The drive signal produced by PM-1 has a carrier frequency of $\Omega_1$ with a modulation depth of $\beta_1$, and is phase modulated by PM-2 at $\Omega_2$ with a modulation depth of $\beta_2$. Therefore, the electronic field of the laser exiting the EOPM-2 can be written as:

$$E_{ESB} = E_0 exp\left\{i\left[\omega t + \beta_1 sin\left(\Omega_1 t + \beta_2 sin\Omega_2 t\right)\right]\right\} \tag{1}$$

Where $E_0$ is the amplitude incident on the EOPM-2, $\omega$ is the carrier frequency of the incoming laser. Using Bessel functions and ignoring the contribution of high order items, we can expand this expression to first order in $\beta_i$ ($i=1, 2$), respectively, as follows:

$$\begin{aligned} E_{ESB} &\approx E_0\left[J_0(\beta_1) + 2iJ_1(\beta_1)sin(\Omega_1 t + \beta_2 sin\Omega_2 t)\right]e^{i\omega t} \\ &\approx E_0 J_0(\beta_1)e^{i\omega t} + E_0 J_1(\beta_1)\left[J_0(\beta_2)e^{i(\omega+\Omega_1)t} + J_1(\beta_2)e^{i(\omega+\Omega_1+\Omega_2)t} - J_1(\beta_2)e^{i(\omega+\Omega_1-\Omega_2)t}\right] \\ &- E_0 J_1(\beta_1)\left[J_0(\beta_2)e^{i(\omega-\Omega_1)t} - J_1(\beta_2)e^{i(\omega-\Omega_1+\Omega_2)t} + J_1(\beta_2)e^{i(\omega-\Omega_1-\Omega_2)t}\right] \end{aligned} \tag{2}$$

Here, $J_n(\beta)$ is the n-order Bessel function. The result of the cascaded phase modulation is to split the light into seven different frequency components: a carrier at $\omega$, two sidebands at $\omega\pm\Omega_1$, and four sub-sidebands at $\omega\pm\Omega_1\pm\Omega_2$, as shown in Fig. 3. The sign of two frequency components at $\omega+\Omega_1\pm\Omega_2$ or $\omega-\Omega_1\pm\Omega_2$ is opposite. Therefore, when the sideband at $\omega+\Omega_1$ or $\omega-\Omega_1$ is locked to a fixed reference cavity, and the carrier and other sidebands are reflected, an ESB error signal can be obtained by demodulating the detected signal with modulation frequency at $\Omega_2$. The frequency discriminant of the ESB locking technique can be given by

$$D_{ESB} = \frac{16FLE_0^2}{c} J_0(\beta_2)J_1(\beta_2)J_1^2(\beta_1) \tag{3}$$

Where $c$ is the speed of light in vacuum, and $L$ is the length of the optical cavity. The finesse is defined as $F = FSR/\nu_{FWHM}$. Here, $\nu_{FWHM}$ is the full width at half maximum of the cavity signal. When one of the sideband is near resonance with the cavity, such as $\omega + \Omega_1 = 2\pi n \cdot FSR$, the carrier frequency of the laser can be tuned by adjusting $\Omega_1$.

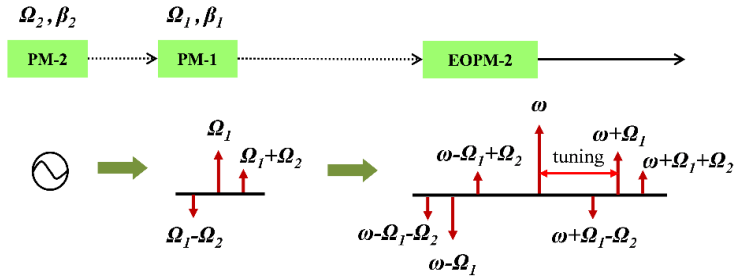

Fig. 3. Modulation structures for the ESB locking technique.

In order to drive the Rydberg transitions of cesium atoms, we use a fiber-coupled waveguide-type EOPM-2 to offset-lock the carrier frequency of the 637.2 nm laser from that of the cavity mode. When a tunable sideband is locked to the ULE cavity, the offset between the objective frequency and the reference frequency depends on a frequency $\Omega_1/2\pi$ with a RF power of 14 dBm. The driven frequency of the EOPM-2 with an RF bandwidth of 5 GHz is phase modulated at $\Omega_2/2\pi$ to produce a couple of modulation sidebands required for ESB locking. The PD-2 output signal is demodulated with the $\Omega_2$ generated by PM-2 (Agilent, 33250A) via phase shift. To optimize the error signal, $\Omega_2/2\pi$ is set to 2 MHz with a modulation amplitude of 10 dBm. The ESB error signal output from the mixer goes into a proportional-integral-differential (PID) controller after a 1.9 MHz low-pass filter (LPF), and then is feedback to the PZT of 1076.9 nm YDFL to obtain a frequency-stabilized tunable 637.2 nm red light. Transmitted signal of the phase-modulated 637.2 nm laser incident on the ULE cavity and the ESB error signal are shown in Fig. 4.

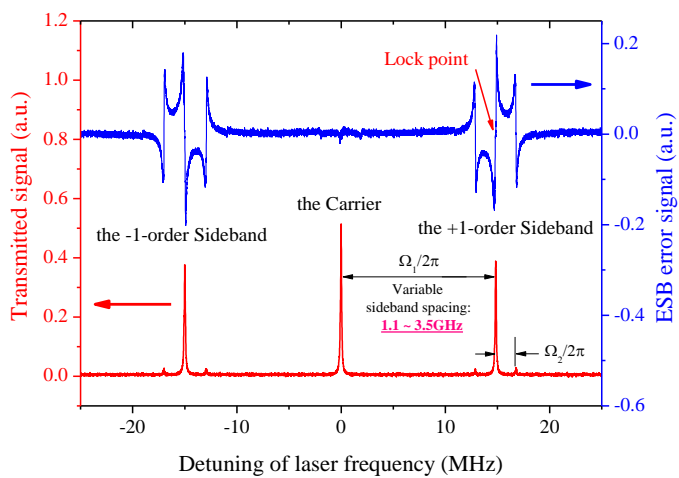

Fig. 4. Transmitted signal of the phase-modulated 637.2 nm laser (red curve) incident on the cavity is obtained by sweeping the carrier frequency of the 1076.9 nm laser while the 1560.5 nm laser remains locked. Blue curve represents the corresponding ESB error signal. Where $\Omega_1/2\pi$ and $\Omega_2/2\pi$ are equal to 15 MHz and 2 MHz with a RF power consumption of 14 dBm and 10 dBm, respectively.

The phase modulation index of the EOPM mainly depends on the applied modulation frequency $\Omega_1/2\pi$, even if the modulation amplitude is constant, as shown in Fig. 5. The modulation index decreases dramatically with the modulation frequency when it is less than 1.1 GHz. Therefore, during sweeping the frequency, the gain of the error signal feedback to the PZT of the 1076.9 nm laser is reduced to prevent the 637.2 nm laser frequency from being locked. To obtain a frequency-stabilized tunable 318.6 nm UV laser, the tuning range of the carrier frequency of 637.2 nm laser is chosen between 1.1 GHz and 3.5 GHz in the gain-flattening region for the modulation index of the EOPM. Where the upper bound of the lasers' tuning range is limited by the 3.5 GHz piezo tuning range of the 1076.9 nm laser.

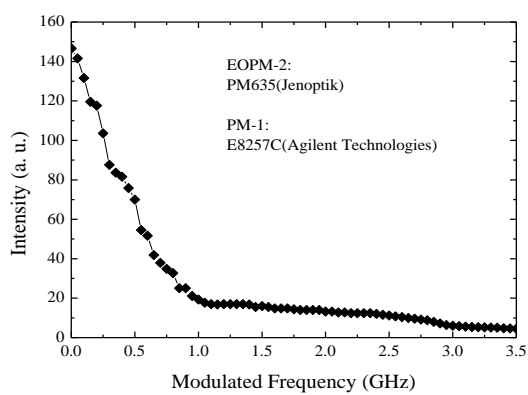

Fig. 5. The phase modulation index of the EOPM-2 depends on the applied modulation frequency.

When the carrier frequency of the 1560.5 nm laser and the upper sideband of the 637.2 nm red light are locked to the ULE cavity, the carrier frequency of the 637.2 nm laser can be tuned by changing the phase-modulated frequency $\Omega_1/2\pi$ produced by PM-1 (Agilent, E8257C). The frequency $\Omega_1/2\pi$ is automatically swept across ~2.4 GHz from 1.1 GHz to 3.5 GHz. The slew rate of the frequency tuning is set to 2.4 MHz/s while the whole laser system remains locked. The sweep frequency is about 1 mHz for the PZT of the 1076.9 nm seed fiber laser. These parameter settings are accomplished by utilizing internal automatic sweep function of PM-1. In experiment, the bandwidth of feedback loops is approximately 20 kHz. So the laser remains locked within the capture range of the ESB lock for the current slew rate. The maximum slew rate of the frequency tuning is mainly limited by the bandwidth of feedback loops to track the requested frequency changes. The tuning range is characterized by a monitor cavity with a FSR of ~487 MHz. The 637.2 nm laser can be swept across more than 4FSRs of the monitor cavity with maintaining lock. Meanwhile, the 318.6 nm UV laser is swept across over 8FSRs of the monitor cavity with a FSR of ~500 MHz in the same conditions, indicating that the continuously tuning ranges of the frequency- stabilized 637.2 nm and 318.6 nm lasers are at least 1.95 GHz and 4 GHz, respectively, as shown in Fig. 6. The tuning range mainly depends on the RF bandwidth and gain-flattening region of the EOPM, and the PZT's tunable range of 1076.9 nm YDFL.

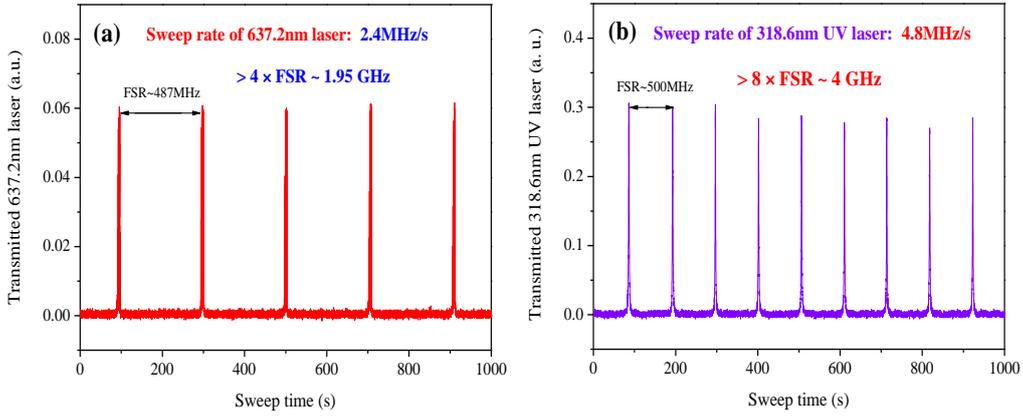

Fig. 6. By changing the upper sideband of the 637.2 nm laser after being locked to the ULE cavity, (a) the tuning range of carrier frequency of the 637.2 nm laser is characterized by an optical cavity with a FSR of ~487 MHz; (b) frequency tuning range of the 318.6 nm UV laser is monitored by a confocal F-P cavity (FSR ~500 MHz; Finesse ~73).

The ULE cavity length is easily affected by the environmental temperature fluctuation, leading to the changes of its resonant frequency, which is associated with the following equation:

$$\alpha \Delta T = \frac{\Delta L}{L} = \frac{\Delta \nu}{\nu} \qquad (4)$$

Where $\Delta T$, $\Delta L$ and $\Delta \nu$ are temperature fluctuation, variation of cavity length, and frequency fluctuation, respectively; $\alpha$ is the material coefficient of thermal expansion, which is related to the material types. The ULE cavity with the coefficient of thermal expansion of $3\times 10^{-8}$/°C is temperature stabilized by a digital temperature controller (Wavelength Electronics, Model LFI-3751). The short-term stability in one hour is better than ±5 mK. Consequently, when the external temperature fluctuation is 0.01 °C, corresponding to variation of ULE cavity length of ~$1.5\times 10^{-11}$ m, leading to the resonance frequency fluctuations are approximately 60 kHz and 140 kHz for 1560.5 nm and 637.2 nm lasers, respectively.

The frequency fluctuations of the free-running 1560.5 nm and 637.2 nm lasers monitored by the ULE cavity are ~8 MHz and ~22 MHz in 30 minutes, respectively. Taking the frequency difference between sidebands and sub-sidebands of the transmission spectrum as a reference frequency ruler, we lock the upper sideband of the carrier frequency to the zero-crossing point, as shown in Fig. 4. The relative frequency stability is estimated using the slope of the zero-crossing point of the ESB error signal. A 2000 s time trace of the ESB error signal converted from voltage to frequency units is shown in the inset of Fig. 7. The relative frequency fluctuation of the 637.2 nm laser after being locked is less than 16 kHz. Fig. 7 shows the relative frequency instability of the red laser locked to the ULE cavity using a servo loop to feedback the 1076.9 nm laser's PZT. We could find the relative frequency instability of the red light using ESB locking technique is less than $1.5\times10^{-11}$ for all recording interrogation time, corresponding to the frequency fluctuation of ~7 kHz. Therefore, the relative frequency fluctuation of the 318.6 nm UV laser is estimated less than 20 kHz. It should be emphasized that the relative Allan deviation is derived from the error signal. The measurement is relatively insensitive to frequency variations introduced by the tiny variation of cavity length, so it only represents a lower limit of the frequency instability. The Allan deviation should be further measured by the beat note of two identical frequency-stabilized laser systems or the optical frequency comb technology, which will more accurately reflect the laser frequency stability. To improve the long-term stability, we can employ EOPM and AOMs [19, 21] for closed-loop frequency feedback corrections and open-loop frequency feedforward corrections in the future.

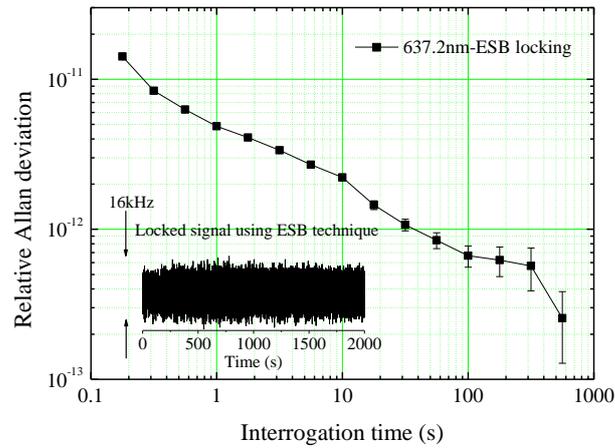

Fig. 7. Relative Allan standard deviation plots show the relative frequency instability of 637.2 nm laser using ESB (black squares) locking technique. Inset is a time trace of the ESB error signal while the 637.2 nm laser is offset locked.

The typical lifetime of cesium $nP_{3/2}$ (n=70-100) state is about several hundred μs [11], the corresponding transition linewidth is estimated approximately a few kHz. Consequently, to drive the single-photon Rydberg transition of cesium atoms for $6S_{1/2}$ - nP (n=70-100), the 318.6 nm UV laser must have narrow linewidth. We use two narrow-linewidth fiber lasers with a linewidth of ~600 Hz for 1560.5 nm DFB-EDFL and ~2 kHz for 1076.9 nm DFB-YDFL, respectively, which are measured by the fiber-delayed self-heterodyne scheme. Moreover, both fiber amplifiers can keep the narrow linewidth. The 637.2 nm laser linewidth is estimated < 3 kHz. Thus, we estimate the linewidth of 318.6 nm UV laser by SHG from 637.2 nm should be < 6 kHz which can be used for the subsequent experimental research.

**4. Conclusion**

In conclusion, we demonstrated a continuously tunable frequency-stabilized 318.6 nm UV laser system. A high-finesse ULE cavity inside a temperature-stabilized ultra-high vacuum chamber is used as frequency reference to stabilize two seed fiber lasers at 1560.5 and 1076.9 nm. The frequency-stabilized 637.2 nm laser by SFG process from two infrared lasers can be widely tuned using ESB locking technique based on a commercial wideband waveguide-type EOPM. Meanwhile, the UV laser can be continuously tuned over 4 GHz with high frequency stability. Further improvement of the tuning ability can be achieved by increasing the RF bandwidth of EOPM, using the fiber laser with widely tunable PZT, and designing the automatic gain control circuit to adjust parameters of servo loops. The narrow-linewidth stable 318.6nm UV laser system with a tunable central frequency is important for the cesium Rydberg experiments, including single-photon Rydberg excitation of cesium atoms, high precision spectroscopy of the Rydberg atoms, and their interactions.

**Acknowledgments**

This project is supported by the National Natural Science Foundation of China (61475091, 11274213, and 61227902) and the National Major Scientific Research Program of China (2012CB921601).